\documentclass[conference]{IEEEtran}
\IEEEoverridecommandlockouts
\usepackage[backend=biber,style=ieee]{biblatex}
\usepackage{hyperref}
\usepackage{adjustbox}
\usepackage{amsmath,amssymb,amsfonts}
\usepackage{graphicx}
\usepackage{multirow}
\usepackage{textcomp}
\usepackage{xcolor}
\usepackage{algorithm}
\usepackage{algpseudocode}
\usepackage{tikz}
\hypersetup{hidelinks}
\usetikzlibrary{arrows.meta, positioning, fit, backgrounds}
\def\BibTeX{{\rm B\kern-.05em{\sc i\kern-.025em b}\kern-.08em
    T\kern-.1667em\lower.7ex\hbox{E}\kern-.125emX}}
\begin{document}

\newcommand\todo[1]{\textcolor{red}{#1}}

\newcommand{\blfootnote}[1]{%
  \begingroup
  \renewcommand{\thefootnote}{}%
  \footnote{#1}%
  \addtocounter{footnote}{-1}%
  \endgroup
}

\title{Toward Composable Network Digital Twins: A Subgraph-Based Latency Prediction Study\\
}

 \author{\IEEEauthorblockN{Shenjia Ding}
 \IEEEauthorblockA{\textit{School of Computing Science} \\
 \textit{University of Glasgow}\\
 Glasgow, UK \\
 2788155d@student.gla.ac.uk}
 \and
 \IEEEauthorblockN{David Flynn}
 \IEEEauthorblockA{\textit{School of Engineering} \\
 \textit{University of Glasgow}\\
 Glasgow, UK \\
 david.flynn@glasgow.ac.uk}
 \and
 \IEEEauthorblockN{Paul Harvey}
 \IEEEauthorblockA{\textit{School of Computing Science} \\
 \textit{University of Glasgow}\\
 Glasgow, UK \\
 paul.harvey@glasgow.ac.uk}
\and
 \IEEEauthorblockN{Takamichi Miyata}
 \IEEEauthorblockA{\textit{Faculty of Advanced Engineering} \\
 \textit{Chiba Institute of Technology}\\
 Chiba, Japan \\
 takamichi.miyata@it-chiba.ac.jp}
\and
 \IEEEauthorblockN{Sumiko Miyata}
 \IEEEauthorblockA{\textit{School of Engineering} \\
 \textit{Institute of Science Tokyo}\\
 Tokyo, Japan \\
 sumiko@ict.eng.isct.ac.jp}

 }

\maketitle

\begin{abstract}

Modern networks must support changing topologies, configurations, and performance objectives, motivating fast and reliable performance estimation. Network digital twins (NDTs) enable what-if analysis for performance estimation in such network scenarios, however, existing machine learning-based NDT approaches often rely on entire topology representations,  which are 
inherently monolithic and lack reusability 
under topological or traffic changes in the network.

This paper introduces a composable NDT approach that decomposes networks into subgraphs represented by reusable \textit{unit twins} that capture subgraph structure, configuration and traffic behaviours.
A lightweight composer aggregates unit twin combinations to create NDTs that predict per-route end-to-end latency through an overall topology. 
Evaluation across controlled synthetic topologies and diverse traffic scenarios, real-world Topology Zoo topologies, and a public NDT challenge dataset demonstrates 
that the composable NDTs achieve high in-distribution accuracy 
while remaining stable under out-of-distribution scenarios.
Comparison with monolithic full topology NDTs demonstrates that our composable approach achieves reusability
, while achieving comparable or superior accuracy.

\end{abstract}


\section{Introduction}

Modern communication networks are becoming increasingly complex and dynamic~\cite{NetworkMultiTaskDT}. The growth of programmable networks, enabled by software-defined network, network function virtualization, and cloud-edge infrastructure, has enabled the flexibility to adapt network configurations to changing operational demands, while simultaneously increasing the complexity of network management, making testing more relevant and challenging~\cite{softwareChallenge}. 
Before deploying new network configurations, it is necessary to understand how the network will behave under different conditions. However, testing on real networks can be costly, risky and difficult to repeat~\cite{syed2020challenges}. Current simulation tools, such as ns-3~\cite{riley2010ns}, can provide detailed network behaviour, but large-scale simulation requires significant time and computational effort~\cite{simulationLimit}.


Network Digital Twins (NDTs)~\cite{NDT} have emerged as a promising approach to support network testing and optimisation, allowing network behaviour to be analysed without directly disturbing the real system. In this context, Graph Neural Networks (GNNs)~\cite{gnnbased} have been used as a key enabler to construct data-driven NDTs for various tasks, such as latency prediction and traffic modelling. GNN-based NDTs can approximate expensive simulation and provide computationally efficient prediction for network testing~\cite{GNNtrends}.


Despite this potential, many GNN-based NDTs are still designed in a monolithic manner.
In such designs, a single GNN model represents an entire network topology, requiring training data that spans all configurations and operational conditions for the full target network topology. 
Such models achieve strong performance on scenarios seen during training, however, their generalisation to unseen conditions is often limited, meaning they struggle to accommodate network modifications, which constrains flexibility, adaptability, and reusability in network testing, see Section~\ref{sec:bg_mono_ndt}.


To address this limitation, this paper explores a \textit{composable} view of NDTs (Section~\ref{sec:methodology}). 
Instead of treating an NDT as a monolithic network predictor, 
we consider an NDT to be a composition of reusable subgraph predictors known as \textit{unit twins} (UTs), which can be trained, stored, and recomposed to create subsequent NDTs. 
Target network topologies are decomposed based on known subgraph shapes, translated into UTs, and then composed to create an NDT of the target network. 
Should the existing set of UTs be insufficient for the task, new UT families can be added without requiring full model retraining.

We evaluate the effectiveness of our approach across synthetic baseline topologies, topology modification, several traffic delay models, a GNNet NDT challenge dataset~\cite{gnnetChallege}, and real-world topologies from Topology Zoo~\cite{topologyzoo} (Section~\ref{sec:results}). 
Results demonstrate that NDTs composed from UTs can effectively predict latency under both seen and unseen topologies and traffic conditions, whereas a centralised GNN-based NDT performs well for seen scenarios, but is less able to generalise to unseen scenarios without retraining.





To the best of our knowledge, this is the first work to study NDTs composed of GNN-based UTs for route-level latency prediction. 
Our focus is not composability as a general digital twin concept, but how subgraph predictors (UTs) can be reused for route-level NDT prediction under network modifications to enable more flexible NDTs.

Our contributions are:
\begin{itemize}
    \item introducing a composable view of NDTs, where subgraph encoders (UTs)  are treated as reusable components for latency prediction. 
    \item formulating subgraph encoder reuse as a graph composition problem via GNN-based latency predictors. 
    \item demonstrating that a GNN-based subgraph composition approach can improve robustness under unseen topology composition and topology modifications compared with centralised GNN models.
\end{itemize}



\section{Background and Motivation}




\subsection{Network Digital Twin}


Network digital twins (NDTs)~\cite{NDT} provide data-driven or simulation-based representations of communication networks for performance monitoring, testing, and control. A key role of an NDT is to support repeated what-if evaluation before changes are deployed to the network by estimating key performance metrics~\cite{Tran2025NetworkDT,Lin20226GDT}, including quality of service (QoS), under varying topology, routing, and traffic conditions~\cite{NDTforPerform}. Latency prediction is especially important as high network latency directly affects user-perceived service quality and the operation of delay-sensitive applications, such as cloud services, video conferencing, industrial control, and edge computing~\cite{LatencyInNDT}.

\subsubsection{NDT Representation}



Simulation-based NDTs~\cite{11197454} can provide detailed performance measurements and remain important when high-fidelity behaviour is required. 
However, repeated simulation over many candidate configurations, topologies, or traffic behaviours can be computationally expensive, which limits their use for rapid network planning and configuration exploration~\cite{ding2026automated}.

Machine learning (ML)-based NDTs provide a complementary approach~\cite{NDT}. 
Wei et al.~\cite{DataDrivenRouting} use deep learning and reinforcement learning to build data-driven network predictions for routing optimisation, aiming to improve simulation speed and accuracy. 
For QoS prediction, Saravanan et al.~\cite{DTtoImproveQoS} use learning models and ensemble methods to predict network delay and optimise service performance. 
Once trained, ML models can act as computationally efficient NDT predictors, estimating network performance without rerunning a full simulator for every candidate configuration~\cite{rusek2020routenet}. 
This makes them attractive for rapid what-if analysis, configuration exploration, and performance-aware network planning.


\subsubsection{Generalizability of Monolithic NDTs}
\label{sec:bg_mono_ndt}

Many existing ML-based NDTs adopt a monolithic design, training a single model to represent an entire network topology~\cite{mozo2022b5gemini,hui2022digital}. While accurate under previously seen conditions, such models struggle when topology or traffic distributions change substantially, as the learned mapping from graph structure to performance no longer holds, requiring retraining on new data spanning the modified topology or context to recover accuracy~\cite{Wang2017MachineLF}. This incurs significant data and computational costs, limiting flexibility, adaptability, and reusability.




\subsection{Digital Twin Composition}
\label{sub:dt-compose}

In digital twins (DTs) more broadly, composition has been proposed to enable flexibility, adaptability, and reuse. Instead of building a single monolithic twin for every target system, a composable DT treats the system as an assembly of smaller models, services, or tools that can be connected through defined interfaces~\cite{gil2025architecture}. This view is especially relevant for complex systems whose components may be added, removed, upgraded, or recombined over time. Existing works can be grouped into two categories:

\subsubsection{Co-Simulation}
approaches connect multiple simulators or behavioural models so that different parts of a networked system can exchange data and jointly evaluate system-level behaviour~\cite{ZakiHindi2026ARF}. This enables DTs to combine physical models, controllers, and external data sources~\cite{frasheri2021rmqfmu}, often via standards, such as the Functional Mock-up Interface (FMI)~\cite{junghanns2021functional}. 
Co-simulation-based DT architectures have been proposed to connect heterogeneous models and support coupled behaviour in complex systems~\cite{gil2025architecture}. In communication networks, CAVIAR~\cite{borges2024caviar} provides a modular co-simulation methodology for 6G scenarios by combining communication simulation, 3D scenes, and AI modules within a shared DT workflow.


A key limitation of this approach is computational efficiency. 
The overall runtime depends not only on the per-component speed, but also on the cost of coordinating their interactions, where repeated data exchange, time synchronisation, and coordination with external simulators or processes are required~\cite{oest2023coupling}. Consequently, even if some components are efficient, the coupled system can be slowed by communication overhead or by the slowest component in the workflow. This limits the suitability of co-simulation for NDT scenarios that require repeated evaluation.

\subsubsection{Modular Architectures}

Several works propose architectural and design principles for composable DTs, noting that 
DT development should support reuse of effort, accelerated time to results, and generalisability~\cite{vanSchalkwyk2023composable}. 
Composite DT architectures describe how multiple DTs can be interconnected to form a collaborative digital ecosystem~\cite{kuruppuarachchi2022architecture}, while DT connectivity studies show that larger DTs may be organised as connected or aggregated structures built from multiple interacting twins~\cite{SCHROEDER2021737}. 
However, most modular DT architectures remain at the level of system design, service composition, governance, or platform integration. While these concerns are essential considerations for DTs, they do not define how the internal units of a DT should be represented or reused, nor do they provide concrete implementations for reusable predictive units in NDTs.

\subsubsection{Summary}
Overall, existing DT-composition research shows a clear movement away from purely monolithic twins. However, most works treat composition as an integration, orchestration, privacy, or interoperability problem. Less attention has been given to composability as a modelling principle for NDTs.

\subsection{Graph Neural Networks}
\label{sec:gnn}

Graph neural networks (GNNs) are deep learning models that learn representations from graph-structured data by propagating and aggregating information across neighbouring nodes and edges. 

\subsubsection{GNN-based Network Representation}


GNNs are well suited to communication network modelling because network topology, routing paths, link attributes, and traffic conditions can naturally be represented as graph features~\cite{GNNforNetwork}. 

For NDTs, GNNs can represent network state and structural dependencies under changing routing, traffic, and configuration conditions.
Prior work has applied GNNs to network performance modelling, routing-aware prediction, and NDT-supported network management~\cite{rusek2020routenet,mozo2022b5gemini,hui2022digital}.
For network performance prediction, GNNs have been used to estimate metrics, such as delay, loss, throughput, and flow-level QoS~\cite{GNNforNetwork}.
Route-level latency prediction is concerned with a source--destination path whose delay depends on route structure, link conditions, queuing state, and neighbouring traffic interactions~\cite{GCNforLatency}.

Many GNN-based network prediction models are trained as full-topology or scenario-level models.
They map a complete graph, route context, and configuration features to a performance prediction, and are effective when deployment topologies are close to previously seen examples.
When a network is modified, extended, or recomposed, the prediction accuracy often decreases, requiring retraining, as discussed in Section~\ref{sec:bg_mono_ndt}. 

\subsubsection{Distributed, Subgraph-Based, and Modular GNNs}

Beyond full-topology prediction, research on GNNs has studied graph partitioning, subgraph processing, and compositional representations.
As full-graph GNN training can become expensive on large graphs, 
distributed GNN systems can be used to address this by partitioning the graph across machines or devices.
For example, DistGNN~\cite{md2021distgnn} uses graph partitioning and communication reduction for scalable full-batch GNN training. 
ByteGNN~\cite{zheng2022bytegnn} introduces GNN-aware partitioning and scheduling to reduce communication overhead in distributed training. 
Compositional graph models also view graph representations as being built from parts or substructures~\cite{kondor2018covariant}. 
Subgraph-level prediction methods, such as SLAPP~\cite{wang2024slapp}, show that subgraphs can support larger-system prediction, although not for communication network latency.
Similarly, graph pooling and subgraph representation methods suggest that subgraph-level prediction can benefit from structured intermediate representations~\cite{Zhang2019HierarchicalGP}.

Our work is inspired by these partitioned and compositional approaches, but uses them for a different purpose: 
graph decomposition is used for reusable unit twins in NDT construction, as opposed to distributed GNN training.

\begin{figure*}
    \centering
    \includegraphics[width=0.8\textwidth]{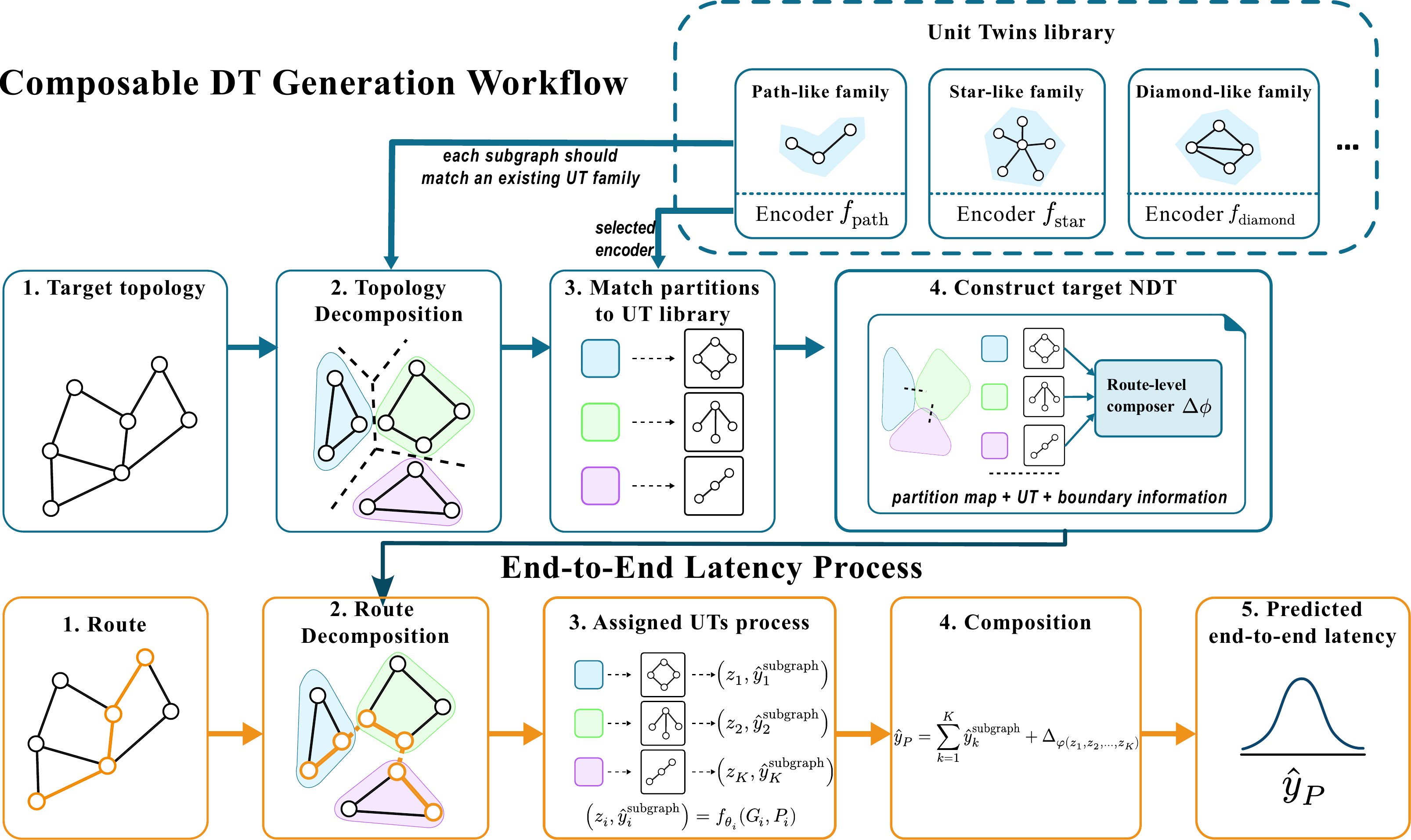}
    \caption{Composable NDT workflow. \textcolor{blue}{Blue} boxes show the process to create an NDT. \textcolor{orange}{Orange} boxes show the process to use an NDT for latency prediction. }
    \label{fig:workflow}
\end{figure*}

\section{Methodology: Composable NDTs}
\label{sec:methodology}






To address the challenge of flexible NDTs, we now present a composable NDT approach for route-level latency prediction under network modifications. 

\subsection{Concept in Nutshell}
As shown in Figure~\ref{fig:workflow}, we use pre-defined network topology \textit{shapes} to decompose a target network topology into subgraphs and then train GNN-based representations of these subgraphs to create reusable \textit{unit twins} (UTs), see Section~\ref{sec:method_unit_twin}. We then compose these UTs into NDTs via a \textit{composer}, see Section~\ref{sec:method_composer}. 

Given a route through the target network, as well as the configuration, 
the latency is predicted by decomposing the route into per-UT segments, predicting the latency of each segment, and combining to obtain the end-to-end latency.

\subsection{Problem Formulation}

We represent a communication network as a directed graph
\(
G=(V,E)
\)
where \(V\) is the set of nodes and \(E\) is the set of directed links. 
Each node \(v\in V\) has a feature vector \(x_v\) that includes queue-related attributes, such as packet buffer queue size. Each edge \(e=(u,v)\in E\) has a feature vector \(a_e\) for link information, indicating bandwidth, capacity, utilisation, traffic load, or route-membership indicators. 
A queried route \(P\) is represented as an ordered sequence of nodes \((v_0,v_1,\ldots,v_H)\), where we assume that each pair of consecutive network nodes is connected by a single link in this work.

End-to-end latency prediction is route-aware: the ordered path \(P\), 
its directed edges, and route descriptors (see Section~\ref{sec:method_route—decom}) are part of the input.
The objective is to learn a predictor \(\hat{y}_P = F(G,P)\) for the end-to-end latency of a route through $G$
but with a composable structure. Instead of training a monolithic predictor over the full graph \(G\), we train predictors based on subgraphs, predict the per-subgraph latency, and compose the values to get the end-to-end latency. 



\subsection{Unit Twin and UT Library}
\label{sec:method_unit_twin}


A \emph{unit twin} (UT) is the basic reusable component of the proposed composable NDT. It represents a subgraph network region and provides a per-route latency-prediction function through the region. In this work, each UT is realised by a pre-defined network topology family and a route-aware GNN encoder. For a subgraph \(G_i=(V_i,E_i)\) and a route segment \(P_i\subseteq G_i\), the UT produces
\begin{equation}
    (z_i,\hat{y}^{subgraph}_i)=f_{\theta_i}(G_i,P_i),
    \label{eq:ut output}
\end{equation}
where \(z_i\) is a learned representation of the subgraph-level route segment and \(\hat{y}^{subgraph}_i\) is the predicted latency contribution.
The pair \((z_i,\hat{y}^{subgraph}_i)\) is the standardized UT output. This allows different UTs to be composed in a uniform way: the route-level composer only consumes the ordered sequence of UT outputs, rather than the internal details of each encoder.

A UT is associated with a network topology family, rather than a single fixed graph instance. 
Currently, the UT library contains five families: \textit{path-like}, \textit{star-like}, \textit{ring-like}, \textit{diamond-like}, and \textit{small-block} regions. They represent structurally distinct local connectivity patterns, including linear forwarding, hub-based branching, cyclic connectivity, and multiple alternative paths. 
These families are selected as representative subgraphs for controlled composition experiments and for the decomposed regions used in our evaluation. They are not intended to cover all possible network subgraphs, but can be extended.


\subsection{GNN Realisation of Unit Twins}
\label{sec:method_gnn_unit_twin}


For a family \(c\), the corresponding route-aware GNN encoder is trained on subgraphs whose node counts lie within a predefined range. 
This allows processing of variable-size subgraphs from the same family, rather than only one fixed node configuration.
In this work, we use a range of three to eight; however, this is configurable.

Node and edge features encode subgraph state, link attributes, and route descriptors (see Section~\ref{sec:method_route—decom}). 
Route descriptors are included to make the encoder route-aware: as the same subgraph may be reused for different route segments, the route must be represented as part of the input rather than being fixed by the topology.
Rather than training a dedicated GNN per matched subgraph instance, we train one shared route-aware encoder for each UT family. 


This design is inspired by distributed and subgraph-based GNN training, where large graphs are divided into smaller regions so that local message passing and representation learning can be performed more efficiently~\cite{md2021distgnn}. 
Our use of decomposition is different: distributed GNN methods mainly partition graphs to scale the training of one global model, while we use subgraph families to define reusable predictive units for NDT construction.

For each UT family \(c\), the encoder \(f_{\theta_c}\) maps a subgraph and route segment to a standardised output as in Eq.~\ref{eq:ut output}.
The encoder consists of message passing, route-aware readout, and two output heads:
\begin{equation}
    h_i = \mathrm{Readout}_{P_i}\big(\mathrm{GNN}_{\theta_{c_i}}(G_i,P_i)\big),
    \label{eq:readout_gnn}
\end{equation}
\begin{equation}
    z_i = g^{emb}_{\theta_{c_i}}(h_i),
\qquad
\hat{y}^{subgraph}_i = g^{pred}_{\theta_{c_i}}(h_i).
    \label{eq:output_heads}
\end{equation}
Message passing updates node and edge representations within \(G_i\), while the route-aware readout aggregates the representations associated with the route segment \(P_i\). 
The embedding head produces \(z_i\) for composition, and the prediction head estimates the subgraph latency contribution. 
As all UT encoders expose the same output format, heterogeneous subgraphs can be reused and recomposed for end-to-end latency prediction.
The architecture of the UT encoder that produces \(z_k\) and \(\hat{y}^{subgraph}_k\) is described in Section~\ref{sec:exp_ut_encoders}.

As message passing is performed on subgraphs and trained encoders can be reused across matched UT instances, our approach reduces the need to train a new monolithic GNN for every target topology. 


\subsection{Topology Decomposition and Route Decomposition}
\label{sec:method_composable_ndt}

We now describe how to take a target network topology and represent it as UTs, as well as how to predict the end-to-end latency for a route through the network. 
The approach has two connected pipelines, as shown in Figure~\ref{fig:workflow}. 




\subsubsection{Topology Decomposition}
is the process of decomposing a target topology into subgraphs based on the pre-defined topology families described above, where each family has a pre-defined GNN encoder (Figure~\ref{fig:workflow}, blue boxes).
Topology mapping produces edge-disjoint subgraphs (Section~\ref{sec:results_controlled} discusses topology decomposition methods)
\(
G_1,\ldots,G_M,
\) where 
each physical link belongs to exactly one subgraph:
\begin{equation}
    E_i \cap E_j = \emptyset, \quad i\neq j,
    \qquad
    E = \bigcup_{i=1}^{M}E_i.
    \label{eq:edge-disjoint}
\end{equation}
The vertex sets may overlap, allowing adjacent UTs to share boundary nodes. 
If a subgraph cannot be matched to an existing UT family, direct reuse is not possible for that region and the library must be extended with a new family.
Across the topologies evaluated in this paper in Section~\ref{sec:datasets}, the five listed UT families cover all partitioned subgraphs after topology decomposition. 

\subsubsection{Route Decomposition}
\label{sec:method_route—decom}
Having generated the UTs and composed them to create an NDT, we can now estimate the latency for a given end-to-end route through the target topology, as shown by the orange boxes in Figure~\ref{fig:workflow}. 
A route is first decomposed into an ordered sequence of subgraph route segments, based on the UTs in the NDT. 
Given a queried route
\(
P=(v_0,v_1,\ldots,v_H),
\)
the route is rewritten, using the decomposed subgraphs, as
an ordered sequence of subgraph route segments:
\begin{equation}
    P = P_1 \oplus P_2 \oplus \cdots \oplus P_K.
    \label{eq:route_segments}
\end{equation}
Each segment \(P_k\) is the maximal consecutive part of the route whose edges belong to the same partitioned subgraph. 
It is processed by the UT assigned to that subgraph.
The route is represented as an ordered UT sequence:
\begin{equation}
    (\mathcal{T}_{s_1},P_1),
    (\mathcal{T}_{s_2},P_2),
    \ldots,
    (\mathcal{T}_{s_K},P_K),
    \label{eq:UT_sequence}
\end{equation}
where \(\mathcal{T}_{s_k}\) is the UT used for segment \(P_k\). 








\subsection{Route-Level Composer}
\label{sec:method_composer}

After route decomposition, 
the route-level composer combines UT outputs to predict the end-to-end latency of the route.

As end-to-end latency \(\hat{y}_{P}\) may include non-additive effects caused by boundary interactions, bottlenecks, or congestion coupling between neighbouring route segments, we therefore add a learned route-level correction term \(\Delta_{\phi}\), which represents the residual adjustment added to the summed subgraph latency prediction:
\begin{equation}
    \hat{y}_{P}
    =
    \sum_{k=1}^{K}\hat{y}^{subgraph}_k
    +
    \Delta_{\phi}(z_1,z_2,\ldots,z_K),
    \label{eq:correction}
\end{equation}
where \(z_k\) is the embedding of the \(k\)-th subgraph segment and \(\hat{y}^{subgraph}_k\) is its predicted subgraph latency contribution. 

In the composed NDT, \(\Delta_{\phi}\) is implemented using a gated recurrent unit (GRU). A GRU is a recurrent neural network module that processes a sequence step by step while maintaining a hidden state. 
As UT outputs are naturally ordered along the queried route, a GRU is well suited to the problem setting. 
Given the ordered embeddings \(z_1,\ldots,z_K\), the GRU updates its hidden state as
\begin{equation}
    h_k=\mathrm{GRU}_{\phi}(z_k,h_{k-1}),
    \label{eq:gru}
\end{equation}
and the final hidden state is mapped to the correction term using an MLP readout.
We refer to the full NDT using UT encoders and a GRU-based composer as the \emph{Composable GRU}. Its architecture is described in Section~\ref{sec:exp_composable_gru}.





\section{Experimental Setup}
\label{sec:experiments}

We evaluate the proposed composable NDT approach through a staged experimental design, moving from controlled synthetic settings to congestion-aware traffic scenarios and 
simulator-generated network traces. 

\subsection{Evaluation Protocol}

\subsubsection{Evaluation Scenarios and Size Definitions}
\label{sec:exp_splits}

We evaluate the accuracy of UT encoders and composer under seen scenarios, 
comprising held-out samples from the same design space as the training data, and unseen scenarios that differ from this design space, as described in
Table~\ref{tab:evaluation_splits}. 



\begin{table}[t]
\centering
\caption{Evaluation splits and their generalisation purpose.}
\label{tab:evaluation_splits}
\resizebox{\linewidth}{!}{
\begin{tabular}{l p{0.45\linewidth} p{0.32\linewidth}}
\hline
\textbf{Scenario} & \textbf{Difference from training set} & \textbf{Purpose} \\
\hline
Seen & Held-out samples from the same design space & In-distribution prediction \\
Unseen composition & Arrangement of UTs in NDT& Compositional generalisation \\
Unseen size & Number of nodes per UT & Size generalisation \\
Unseen composition + size & Number of nodes per UT and UT arrangement in NDT
& Hardest systematic generalisation setting \\
Unseen traffic regime & Arrangement of UTs in NDT and per-UT traffic delay regime & Traffic-behaviour generalisation \\
\hline
\end{tabular}
}
\end{table}

\subsubsection{Training Protocol and Variants}



\paragraph{UT Encoder Training}
\label{sec:exp_ut_encoders}


Each UT family was trained using \(6,000\) route-conditioned samples generated from multiple instances of the corresponding subgraph family, including subgraphs of different sizes where permitted by the family definition. For each subgraph instance, the network configuration was varied by changing parameters, including the source--destination pair, route, bandwidth, and queue size. The latency labels depend on the experiments, as shown in Section~\ref{sec:datasets}.

All family encoders used the same architecture and optimisation settings. The hidden dimension was set to \(64\), with three GINE message-passing layers and a dropout rate of \(0.1\). The encoders were trained for \(200\) epochs using Adam with a learning rate of \(10^{-3}\), weight decay of \(10^{-5}\), and batch size \(128\). A fixed random seed was used.

The encoder produces two outputs.
First, it generates a compact 64-dimensional embedding that represents the subgraph in a way useful for composition.
Second, it predicts a scalar value of the estimated subgraph segment latency.

\paragraph{centralised GNN (baseline)}
is a full-topology reference model. 
It receives the same node features, edge features, and route descriptors as the composable models, but predicts on the complete topology rather than on a sequence of UT outputs. 
The GNN uses three GINE message-passing layers with a hidden dimension of \(64\), residual connections, ReLU activations, global mean pooling, and a dropout rate of \(0.1\). 
The model was trained for \(200\) epochs using Adam with a learning rate of \(10^{-3}\), weight decay of \(10^{-5}\), a batch size of \(64\), and mean-squared error as the training loss. 

For each experiment, it is trained using the same train/validation/test splits and full-route latency labels as the composable models. 
The same optimizer, learning-rate setting, early-stopping criterion, and evaluation metric are used unless otherwise stated.

This baseline represents the standard full-graph learning setting, where the model can observe the full topology and route context directly. It is not designed as a reusable compositional model, but to provide a comparison point for evaluating the  cost and benefit of using composed subgraph predictors.

\paragraph{Encoder Summation (baseline)}
is the no-correction summation baseline. 
It removes the learned correction term in Eq.~\ref{eq:correction} and predicts route latency only by summing the UT latency estimates. 
This baseline isolates the contribution of the learned composer by testing how far subgraph latency predictions can explain end-to-end latency without correction.

\paragraph{Composable GRU}
\label{sec:exp_composable_gru}
is the main sequence-aware variant described in Section~\ref{sec:method_composer}. 
For a route decomposed into \(K\) segments, the corresponding 64-dimensional UT embeddings are arranged according to their traversal order and processed by a single-layer GRU with a hidden dimension of \(64\). The composer consists of a single-layer GRU followed by a two-layer correction head, 
which produces the correction \(\Delta_\phi\), which is added to the summed UT predictions.

\paragraph{Composable MLP} is an order-invariant ablation uses the same UT encoders and latency estimates as the main approach, but computes the correction from mean-pooled UT embeddings.
A multilayer perceptron (MLP) is a feed-forward neural network that maps an input vector to an output prediction through fully connected layers. 
The valid 64-dimensional UT embeddings are mean-pooled to form a fixed-dimensional route representation, while a mask excludes padded segments. 
This representation is processed by a three-layer MLP with 64-dimensional hidden layers, ReLU activations, and a dropout rate of \(0.1\). 
Because mean pooling removes traversal order, this variant tests whether UT predictions plus a route-level correction are sufficient.

\subsubsection{Metrics}
\label{sec:metrics}

We evaluate prediction accuracy using the coefficient of determination, \(R^2\), as the main metric. 
As latency prediction is a regression task and our experiments span different topology sizes, traffic models, and latency ranges, \(R^2\) provides a scale-normalized measure of 
comparison 
across heterogeneous settings.
\(R^2\) values close to 1 indicate accurate prediction. \(R^2=0\) means the performance is no better than predicting the mean latency of the test split. Negative \(R^2\) values indicate worse-than-mean prediction and are useful for identifying failed transfer or poor generalisation under topology changes.

\subsection{Datasets}
\label{sec:datasets}

\subsubsection{Controlled Synthetic Datasets}
\label{sec:dataset_controlled}

We first construct controlled synthetic datasets to study how topology decomposition affects UT composition. 
It is necessary to determine whether UTs can be composed reliably and which decomposition strategy provides a clear basis for reuse. We focus on identifying decomposition strategies that maximise opportunities for UT reuse rather than on finding graph partitions that are theoretically or computationally optimal. Consequently, decomposition quality is assessed in terms of reusability and composability, not partition optimality.

In the synthetic dataset, each dataset sample consists of a composed topology, a route, node and edge features, and a route-level latency label. 
The topologies are generated from compositions of topologies and sizes of the UT families described in Section~\ref{sec:method_unit_twin}.
Node features include route indicators, route position, and queue size; edge features include bandwidth and route-membership descriptors. 
These features are processed during message passing, while route descriptors indicate relevant route segment nodes and links.

\subsubsection{Local Topology Change}
To evaluate reuse when a subset of the full topology changes, we construct topology-change datasets. 
Each data sample contains a base topology and a changed topology. The changed topology randomly modifies a subgraph of the target topology. 
This setting captures the intended reuse scenario of a composable NDT: when a network changes locally, prediction should be maintained by reusing existing UTs, without rebuilding a full topology-level model. It specifically evaluates cases where the modified topology remains expressible within the existing UT library, avoiding the need for full remapping.

\subsubsection{Congestion-Aware Traffic Dataset}
\label{sec:dataset_congestion}
Beyond topology modification, we also consider congestion-aware datasets to understand if UT composition remains effective 
when route latency depends on traffic load, 
rather than topology structure alone. 
We use three analytical delay-generation models to create 
utilization-driven queuing (M/M/1~\cite{bertsekas2021data}), arrival and service variability (Kingman G/G/1~\cite{Kingman_1961}), and bound-style delay targets (network-calculus-inspired delay model~\cite{networkcalculusdelay}) for 
traffic-dependent latencies 
to evaluate NDT composition.
For each link \(e\), utilisation is defined as
\begin{equation}
    \rho_e = \frac{\lambda_e}{\mu_e},
    \label{wq:utilisation}
\end{equation}
where \(\lambda_e\) denotes the average rate at which packets arrive at link \(e\), while \(\mu_e\) denotes the average rate at which packets can be served or transmitted by that link. 

We use three congestion regimes: \textbf{low} \(\rho_e\sim U(0.10,0.30)\), \textbf{medium} \(\rho_e\sim U(0.40,0.65)\), and \textbf{high} \(\rho_e\sim U(0.70,0.90)\). 
The ranges are chosen to create distinguishable traffic regimes while avoiding degenerate cases. Very low utilisation would remove most queuing effects, while values close to one can make queuing-based delays numerically unstable and dominate the learning target. The high regime represents strong but stable congestion pressure rather than near-saturation behaviour.


\subsubsection{GNNet DT Challenge Dataset \& Topology Zoo}
We use two external sources to evaluate whether the proposed approach can operate beyond manually constructed topology compositions.

\paragraph{GNNet NDT Challenge Dataset}
The publicly available NDT challenge GNNet simulation traces~\cite{gnnetChallege} are used 
to evaluate whether the proposed approach can operate beyond controlled analytical latency generation. 
GNNet is a network-performance dataset from an NDT challenge. It was generated from OMNeT++ packet-level network simulation and provides topology, routing, traffic, and route-level delay labels, allowing us to test the approach on data produced by a network simulator. 

The evaluation on the GNNet challenge dataset should be interpreted as a bridge between the synthetic evaluation and broader real-world validation. It shows that the 
reusable UT encoder representations and the 
composer form a viable workflow on OMNeT++-generated network traces. 

\paragraph{Topology Zoo}
To evaluate our approach on real-world network topologies, we use Topology Zoo~\cite{topologyzoo}. 
Topology Zoo is a public collection of network topologies collected from real operator, research, and backbone networks.
Since Topology Zoo provides topology structure but not measured latency labels, we generate route-level latency targets using the approach described in Section~\ref{sec:dataset_controlled}. 
The topology decomposition was manually specified in the current experiments.
For each network, we inspected the graph structure and identified subgraphs in the UT libraries.



\begin{table*}[t]
\centering
\caption{
Comparison of Topology Decomposition \& NDT Composition. Values are \(R^2\).}
\label{tab:controlled_decomposition}
\begin{adjustbox}{max width=\textwidth} 
\begin{tabular}{l l c c c}
\textbf{Decomposition setting}
& \textbf{NDT}
& \textbf{Seen}
& \textbf{Unseen comp.}
& \textbf{Unseen size} \\ \hline
\multirow{3}{*} {Edge-disjoint, single boundary node}
& Composable GRU
& \textbf{0.9995}
& \textbf{0.9912}
& \textbf{0.9633} \\ 
& Composable MLP
& 0.9988
& 0.9799
& 0.7310 \\ 
& Encoder Sum.
& 0.9694
& 0.9217
& 0.3486

\\
\hline
\multirow{3}{*} {Edge-disjoint, multiple boundary nodes}
& Composable GRU
& \textbf{0.9994}
& \textbf{0.9969}
& 0.9155 \\
& Composable MLP
& 0.9986
& 0.9885
& \textbf{0.9475} \\  

& Encoder Sum.
& 0.9366
& 0.9230
& 0.6746
\\

\hline
\multirow{3}{*} {Edge-shared}
& Composable GRU
& \textbf{0.9912}
& -2.8882
& -2.7486 \\
& Composable MLP
& 0.9383
& \textbf{0.8456}
& 0.4963 \\
& Encoder Sum.
& 0.6184
& 0.5944
& -1.0242
\\
Edge-shared with interface information
& Interface-aware composer
& 0.9758
& 0.8211
& \textbf{0.6975} \\
\hline
\end{tabular}
\end{adjustbox}
\end{table*}

\section{Results}
\label{sec:results}

\subsection{Controlled Validation of Unit Twin Composition}
\label{sec:results_controlled}




Table~\ref{tab:controlled_decomposition} compares two topology decomposition designs using the controlled synthetic data:



\paragraph{Edge-Disjoint} assigns each physical link to exactly one UT, while adjacent UTs may share boundary nodes.
Both boundary-node variants achieve high accuracy under all scenarios, showing that pretrained UT encoders from known families can be recomposed for latency predictions, 
suggesting that preserving the ordered sequence of UT outputs is useful for route-level composition.

\paragraph{Edge-Shared} decomposition allows a physical link to appear in more than one UT. 
This diagnostic setting tests whether overlapping link ownership makes subgraph latency attribution less stable. 
Although the results show strong performance on seen configurations, their behaviour is less stable under unseen scenarios, 
suggesting that shared physical links make subgraph latency ownership ambiguous. The same link-level contribution may be represented by multiple UTs, making composed prediction harder to generalise.

An interface-aware composer is included as a diagnostic variant for this setting. 
It augments the composer with additional information about the UT overlaps, testing whether explicit boundary information can compensate. 
The results 
improve stability over the failing GRU case, but do not recover the performance of the edge-disjoint node-overlap design. 
This indicates that the weakness is not simply caused by insufficient composer capacity. 
Rather, duplicating physical links across UTs weakens the interpretation of latency contributions, and makes systematic composition harder under unseen scenarios. 


Based on these results, edge-disjoint is used in the following experiments.

\subsection{Reuse under Topology Modification}

Table~\ref{tab:topology_change_reuse_1000} evaluates whether the composable NDT can remain effective after a target topology is modified.
We consider three scenarios: NDT performance on the original \textit{base} topology, \textit{zero-shot} reuse on the modified topology without retraining, and adaptation using 1000 samples from the modified topology to \textit{retrain} the composer. We consider each scenario under three traffic delay models, where they are applied homogeneously.
Because the objective of this experiment is to evaluate reuse under local topology modification rather than to evaluate different composer designs, we focus on comparing the main composable approach, Composable GRU, with the centralised GNN baseline.


\begin{table}[t]
\centering
\caption{Topology-change reuse results with 1000 changed-topology adaptation samples. Values are \(R^2\).}
\label{tab:topology_change_reuse_1000}
\begin{adjustbox}{max width=\linewidth}
\begin{tabular}{l l c c c}
\hline
\textbf{Delay Model}
& \textbf{NDT}
& \textbf{Base}
& \textbf{Zero-shot}
& \textbf{Retrained} \\
\hline
\multirow{2}{*}{M/M/1}
& Composable
& 0.985
& \textbf{0.982}
& \textbf{0.986} \\
& centralised
& \textbf{0.996}
& -1.064
& 0.952 \\
\hline
\multirow{2}{*}{Kingman G/G/1}
& Composable
& 0.967
& \textbf{0.973}
& \textbf{0.975} \\
& centralised
& \textbf{0.978}
& 0.039
& 0.952 \\
\hline
\multirow{2}{*}{Network calculus}
& Composable
& 0.981
& \textbf{0.981}
& \textbf{0.982} \\
& centralised
& \textbf{0.983}
& -0.626
& 0.963 \\
\hline
\end{tabular}
\end{adjustbox}
\end{table}

The centralised GNN achieves strong performance on the base topology. 
However, its zero-shot performance drops substantially, with negative \(R^2\) for M/M/1 and network-calculus traffic, and close to zero for Kingman G/G/1. 
This indicates that the full-topology representation is sensitive to local structural changes without retraining.
In contrast, the composable NDT zero-shot performance remains close to its base-topology performance across all three traffic models. 
This suggests that the changed topology can still be represented through a new assembly of known UTs, allowing the NDT to remain useful without retraining.

After retraining, the centralised GNN recovers substantially, reaching positive \(R^2\) across all three traffic models. 
However, it remains below the adapted composable NDT in these tests. 
The composable NDT also changes only slightly from zero-shot to adapted performance, suggesting that most of its benefit in this setting comes from UT reuse rather than from retraining. 
This result supports the intended reuse scenario of the composable NDT: when a topology changes locally but remains expressible through known UTs, route-level prediction can be maintained by recombining reusable UT outputs instead of rebuilding a monolithic full-topology predictor.

\subsection{Traffic-Focused Evaluation: Congestion-Aware Traffic}
\label{sec:results_congestion}



Table~\ref{tab:mixed_regime_results} describes the ability of the composable NDT to generalise across network latency scenarios for a fixed topology using the congestion-aware traffic dataset. 

For each traffic delay model, we consider two training settings for the composer: \textit{single-train} setting, the composer is trained on traffic delay data from a single congestion regime and tested on both homogeneous and heterogeneous regimes; \textit{mixed-train}, heterogeneous congestion delay data samples are also included during training. This allows us to distinguish the effect of compositional structure from the effect of exposure to mixed traffic conditions.

\subsubsection{Homogeneous Congestion Regime}
The \textit{Low}, \textit{Medium} and \textit{High} columns in Table~\ref{tab:mixed_regime_results} report on composer accuracy for the corresponding congestion regime. 
In this setting, all UTs within an NDT along a route use the same congestion regime and the same delay-generation model, see Section~\ref{sec:dataset_congestion}. 
This tests whether UT prediction and composition remain accurate under 
different congestion separately. 

\textit{Encoder Sum.} 
consistently underperforms the composable approaches, particularly under the \textit{composition + size} split, showing that UT predictions alone are not sufficient in congestion scenarios.

The centralised GNN performs well in seen and unseen-composition settings. 
However, its performance drops sharply under the composition-plus-size split across all traffic delay models. In contrast, the composable GRU and MLP NDTs remain positive in this split and remain comparable across all categories. 
This suggests that the hardest case is not congestion modelling alone, but congestion-aware prediction under a changed UT composition and route length.

\subsubsection{Heterogeneous Congestion Regime}
In this setting, we consider the \textit{mixed-regimes} scenarios, where UTs along the same route use different congestion regimes from the same delay-generation model. 




\begin{table*}[t]
\centering
\caption{Results across homogeneous and heterogeneous congestion scenarios. Values are \(R^2\).}
\label{tab:mixed_regime_results}
\resizebox{\textwidth}{!}{
\begin{tabular}{l l c c c c c c c}
\hline
\textbf{Delay Model}
& \textbf{Composer}
& \textbf{Low}
& \textbf{Medium}
& \textbf{High}
& \textbf{Mixed-seen}
& \textbf{Mixed-unseen}
& \textbf{Unseen comp.}
& \textbf{Comp. + size} \\
\hline
\multirow{7}{*}{M/M/1}
& Comp. GRU, single-train & 0.983 & 0.988 & 0.987 & 0.983 & 0.983 & 0.980 & 0.917 \\
& Comp. GRU, mixed-train & 0.984 & 0.987 & 0.986 & 0.986 & 0.984 & 0.956 & 0.866 \\
& Comp. MLP, single-train & 0.983 & 0.987 & 0.987 & 0.983 & 0.981 & 0.985 & \textbf{0.921} \\
& Comp. MLP, mixed-train & 0.981 & 0.987 & 0.986 & 0.985 & 0.983 & 0.987 & 0.895 \\
& Central, single-train & \textbf{0.996} & \textbf{0.996} & \textbf{0.993} & 0.977 & 0.979 & 0.978 & -2.533 \\
& Central, mixed-train & 0.991 & \textbf{0.996} & 0.990 & \textbf{0.994} & \textbf{0.993} & \textbf{0.992} & -2.175 \\
& Encoder Sum. & 0.430 & 0.240 & 0.561 & 0.396 & 0.583 & 0.585 & 0.143 \\
\hline
\multirow{7}{*}{Kingman}
& Comp. GRU, single-train & 0.981 & 0.983 & 0.966 & 0.957 & 0.966 & 0.974 & \textbf{0.934} \\
& Comp. GRU, mixed-train & 0.978 & 0.981 & 0.965 & 0.970 & 0.974 & 0.976 & \textbf{0.934} \\
& Comp. MLP, single-train & 0.980 & 0.982 & 0.966 & 0.968 & 0.973 & 0.976 & 0.925 \\
& Comp. MLP, mixed-train & 0.979 & 0.977 & 0.966 & 0.969 & 0.974 & 0.977 & 0.923 \\
& Central, single-train & \textbf{0.989} & \textbf{0.988} & 0.968 & 0.877 & 0.887 & 0.914 & -1.151 \\
& Central, mixed-train & 0.979 & 0.984 & \textbf{0.969} & \textbf{0.974} & \textbf{0.976} & \textbf{0.982} & -0.671 \\
& Encoder Sum. & 0.357 & 0.246 & 0.816 & 0.687 & 0.817 & 0.825 & 0.702 \\
\hline
\multirow{7}{*}{Network calculus}
& Comp. GRU, single-train & 0.973 & 0.983 & \textbf{0.979} & 0.978 & 0.980 & 0.978 & \textbf{0.777} \\
& Comp. GRU, mixed-train & 0.973 & 0.983 & 0.978 & 0.980 & 0.980 & 0.980 & 0.759 \\
& Comp. MLP, single-train & 0.973 & 0.983 & \textbf{0.979} & 0.979 & 0.980 & 0.982 & 0.775 \\
& Comp. MLP, mixed-train & 0.973 & 0.983 & 0.978 & 0.980 & 0.980 & 0.983 & 0.764 \\
& Central, single-train & \textbf{0.982} & 0.985 & \textbf{0.979} & 0.947 & 0.962 & 0.959 & -3.230 \\
& Central, mixed-train & 0.979 & \textbf{0.986} & 0.977 & \textbf{0.982} & \textbf{0.981} & \textbf{0.984} & -4.524 \\
& Encoder Sum. & 0.555 & 0.648 & 0.733 & 0.694 & 0.776 & 0.792 & 0.082 \\
\hline
\end{tabular}
}
\end{table*}


In Table~\ref{tab:mixed_regime_results}, the centralised GNN produces negative \(R^2\) for all scenarios under the \textit{composition + size} setting, while the MLP and GRU composers remain positive. 
This suggests that traffic heterogeneity is not the main source of failure, the harder domain problem arises from topology composition.


We consider single- and mixed-train settings. 
Encoder Sum. performs substantially below the composable approach, confirming that a learned correction term is needed to capture route-level effects 
beyond independent subgraph latency predictions.
For the MLP and GRU composers, mixed training does not consistently improve performance over single-regime training, suggesting that the composer can combine UT outputs from different congestion levels without observing every mixed-level order 
during training. 
For the centralised GNN, mixed training improves some mixed-seen and mixed-unseen cases, but does not resolve the failure under \textit{unseen composition + size}. 
This supports the interpretation that robustness comes from the UT composition structure rather than from simply adding more mixed congestion samples.

Overall, the results show that the GRU and MLP composers maintain positive performance under different traffic scenarios. 

\subsection{Simulator-Trace Evaluation: Challenge Dataset}
\label{sec:results_external}


Table~\ref{tab:gnnet_results} reports the route-delay prediction results on the GNNet NDT training dataset. Composable 
GRU achieves the best overall performance across all three splits, showing that the proposed approach can operate on simulator-generated network traces rather than only analytical data 
and improve prediction through learned correction. 
Centralised GNN also performs strongly, but is worse than Composable GRU under the unseen scenarios.

In contrast, MLP performs much worse, 
suggesting that mean-pooled UT embeddings are insufficient for this dataset.
This indicates that route structure and link-level context are important for capturing topology effects on latency in the GNNet dataset.

\begin{table}[t]
\centering
\caption{Route-delay prediction results on the converted GNNet training dataset. Values are \(R^2\).}
\label{tab:gnnet_results}
\begin{adjustbox}{max width=\linewidth}
\begin{tabular}{l c c c}
\hline
\textbf{NDT}
& \textbf{Val. seen}
& \textbf{Test seen}
& \textbf{Test unseen size} \\
\hline
Composable GRU
& \textbf{0.9893}
& \textbf{0.9881}
& \textbf{0.9772} \\
Composable MLP
& 0.2953
& 0.3484
& 0.2407 \\
centralised GNN
& 0.9852
& 0.9827
& 0.9511 \\
Encoder Sum.
& 0.9870
& 0.9848
& 0.9758 \\
\hline
\end{tabular}
\end{adjustbox}
\end{table}




\subsection{Topology Zoo}
\label{sec:exp_topo_zoo}
Table~\ref{tab:topology_zoo} shows results for six Topology Zoo topologies. Due to space limitation, we only report the first six topologies in alphabetical order from the dataset. 
\emph{Composable GRU} reuses the trained UT encoders and route-level composer trained on the controlled synthetic composition datasets described in Section~\ref{sec:dataset_controlled}. 
\emph{Centralised transfer} baseline applies a monolithic GNN trained on the same controlled synthetic composition datasets to each Topology Zoo graph. 
\emph{Centralised in-topology} trains and evaluates a separate full-topology GNN for each target topology. 
Thus, centralised transfer tests direct cross-topology reuse of a full-graph predictor, while centralised in-topology training tests how well a target-specific full-graph model performs when data from the target topology are available.




\begin{table*}[t]
\centering
\caption{Topology Zoo end-to-end latency prediction results. Values are \[R^2\].}
\label{tab:topology_zoo}
\begin{adjustbox}{max width=\linewidth}
\begin{tabular}{l c c c c c c c}
\hline
\textbf{Topology}
& \textbf{Nodes}
& \textbf{Edges}
& \textbf{Routes}
& \textbf{Comp. GRU}
& \textbf{Cen. transfer}
& \textbf{Cen. in-topology} 
& \textbf{Encoder Sum.}\\
\hline
AARNET
& 19 & 24 & 171
& \textbf{0.998}
& -5.927
& 0.887 
& 0.980
\\
Abilene
& 11 & 14 & 55
& \textbf{0.997}
& -6.589
& 0.061 
& 0.982
\\
AboveNet
& 23 & 31 & 253
& \textbf{0.996}
& -2.313
& 0.940 
& 0.980
\\
ACOnet
& 23 & 31 & 253
& \textbf{0.993}
& -21.570
& 0.804 
& 0.959
\\
AGIS
& 25 & 30 & 300
& \textbf{0.992}
& -1.678
& 0.570 
& 0.975
\\
AI3
& 10 & 9 & 45
& \textbf{0.995}
& -5.530
& 0.419 
& 0.974
\\
\hline
Average
& -- & -- & --
& \textbf{0.995}
& -7.268
& 0.614
& 0.975
\\
\hline
\end{tabular}
\end{adjustbox}
\end{table*}

Across the six evaluated topologies, the composable GRU achieves consistently high \(R^2\), 
with an average of 0.995. This indicates that, when a real topology structure can be decomposed into supported UT families, the trained UT encoders and route-level composer can be reused without training a new full-topology predictor for each graph. 
In this evaluation, the UT families defined in Section~\ref{sec:method_unit_twin} are sufficient to cover all decomposed subgraphs in the Topology Zoo networks.


Compared with the composable GRU, the two centralised baselines highlight the benefit of structural reuse, as 
both consistently perform worse.
The centralised transfer baseline performs poorly on all selected Topology Zoo graphs with negative \(R^2\) values, showing that a monolithic GNN trained on the synthetic composed-topology distribution does not directly generalise to real topology structures. The centralised in-topology model performs better because it is trained using data from each target topology, but its performance is still less stable and remains below the composable NDT in these experiments. 
One possible reason is that the centralised GNN depends on the route diversity available within a single graph, whereas the composable approach reuses UT encoders trained across repeated subgraphs.

Overall, the Topology Zoo results show that the proposed approach can decompose real network topologies into subgraphs represented by supported UT families and reuse learned encoders for latency prediction.


\section{Conclusion}
\label{sec:conclusion}

This paper presents a composable Network Digital Twin (NDT) approach for per-route end-to-end latency prediction based on reusable Unit Twins (UTs) and a route-level composer. Rather than training a single predictor for an entire network, the proposed approach decomposes network topologies into reusable subgraph components that can be retrieved and assembled into NDTs to model new topologies and routes. Across controlled composition experiments, congestion-aware traffic scenarios, topology-change studies, GNNet traces, and Topology Zoo networks, the composable approach achieved high predictive accuracy while maintaining robustness under unseen topology compositions, topology-size changes, heterogeneous traffic-delay conditions, and local network modifications. 

A key finding is that, while centralised GNN-based DTs often achieved comparable or better accuracy in seen settings, their performance deteriorated substantially under the most challenging unseen scenarios, 
in several cases yielding negative $R^2$ values. In contrast, the composable NDT consistently maintained positive predictive performance through the reuse and recombination of previously trained UTs. These results suggest that reusable subgraph-level composition provides a practical and robust mechanism for NDT reuse when networks evolve.

\printbibliography


\end{document}